\documentclass[12pt]{article}
\usepackage{graphicx}
\usepackage{amssymb}
\usepackage{bm}
\evensidemargin \oddsidemargin
\def\0{{\bf 0}}
\def \RC {C\!_{{\scriptscriptstyle R}}}

\def \rU {U_{{\scriptscriptstyle R}}}
\def \U {{\cal U}}
\def \ZM {Z_{{\scriptscriptstyle M}}}

\newcommand{\Bp}{{\bf p}}

\newcommand{\bY}{{\bf Y}}
\newcommand{\bx}{{\bf x}}
\newcommand{\bX}{{\bf X}}
\newcommand{\bz}{{\bf z}}
\newcommand{\bu}{{\bf u}}

\newcommand{\bv}{{\bf v}}
\newcommand{\Ba}{{\boldsymbol \alpha}}
\newcommand{\bb}{{\boldsymbol\beta}}
\newcommand{\Be}{{\boldsymbol\epsilon}}

\newcommand{\bE}{{\bf E}}
\newcommand{\bB}{{\bf B}}
\newcommand{\bA}{{\bf A}}

\def \E {{\cal E}}

\newcommand{\be}{\begin{equation}}
\newcommand{\ee}{\end{equation}}
\newcommand{\ba}{\begin{array}}
\newcommand{\ea}{\end{array}}

\newcommand{\bea}{\begin{eqnarray}}
\newcommand{\eea}{\end{eqnarray}}
\begin{document}

\title{On very short and intense laser-plasma interactions
}


\author{Gaetano Fiore  \\  \\
Dip. di Matematica e Applicazioni, Universit\`a ``Federico II''\\
   V. Claudio 21, 80125 Napoli, Italy;\\         
 I.N.F.N., Sez. di Napoli,
        Complesso MSA, V. Cintia, 80126 Napoli, Italy
}

\date{}

\maketitle

\begin{abstract}
We briefly report on some results regarding the impact of very short and intense laser pulses on a cold, low-density plasma
 initially at rest, and the consequent acceleration of plasma electrons to relativistic energies.
Locally and for short times the pulse can be described by a transverse plane electromagnetic travelling-wave 
and the motion of the electrons  by a purely Magneto-Fluido-Dynamical  (MFD) model
with a very simple dependence on the transverse electromagnetic potential, while the ions can be regarded as at rest; the Lorentz-Maxwell and continuity  equations are reduced to the Hamilton equations of a Hamiltonian system with 1 degree of freedom, in the case of a plasma with constant initial density, or a collection of such systems otherwise.
We can thus describe both the well-known  {\it wakefield} behind the pulse and the recently predicted {\it slingshot effect}, i.e. the backward expulsion of high energy electrons just after the laser pulse has hit the surface of the plasma.


\end{abstract}

\section{Introduction and preliminaries}
\label{intro}

Today  the acceleration of charged particles to relativistic energies has a host of important applications, 
 in particular in:

\begin{enumerate}

\item nuclear medicine, cancer therapy (PET, electron/proton therapy,...);
\item research in structural biology;
\item research in materials science;
\item food sterilization;
\item research in nuclear fusion (inertial fusion);
\item transmutation of nuclear wastes;
\item research in high-energy particle physics.

\end{enumerate}

\begin{figure}
\includegraphics[height=7.8cm]{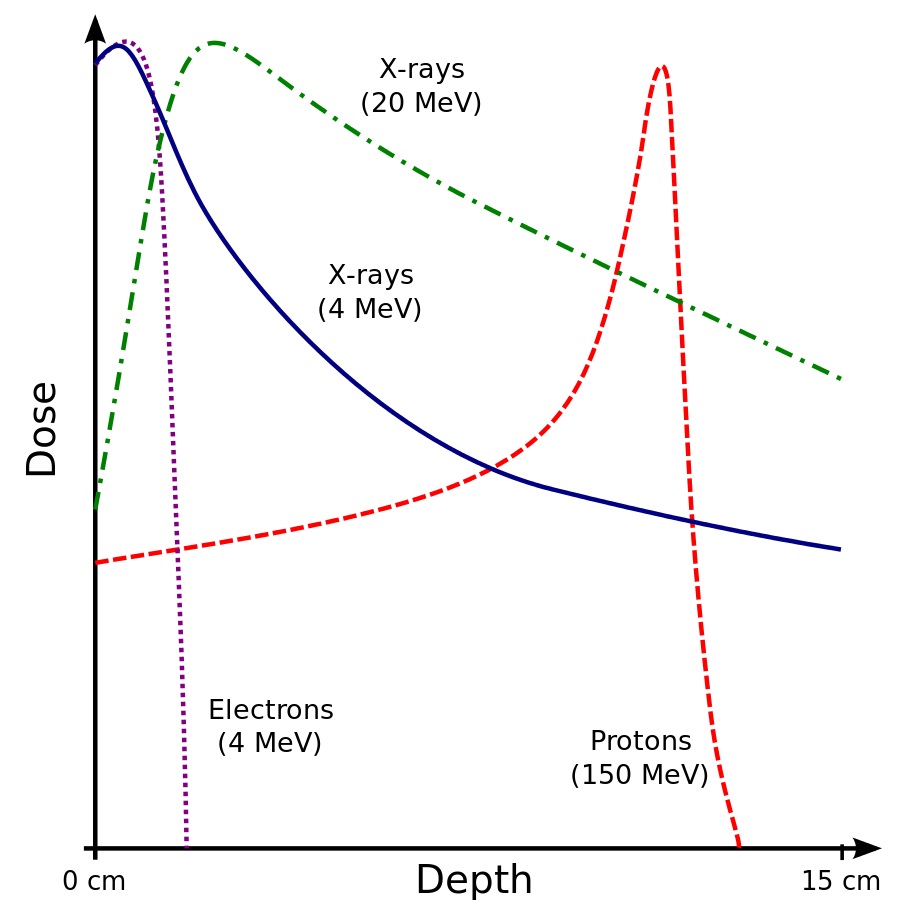}\hfill
\includegraphics[height=7.6cm]{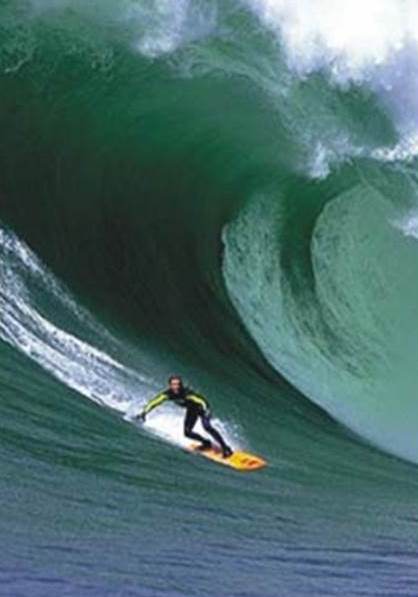}
\caption{Left: rates of absorption of the beam energy by a living tissue (doses) vs. the depth of penetration of the beam.
Right: water wave and surfing.}
\label{Dose_vs_Depth}      
\end{figure}

Let us just mention the main advantage of attacking a cancer by particle rather than by radiation therapy. 
As the dose of X- or gamma-rays absorbed
by  human tissue depends weakly on the depth (see fig. \ref{Dose_vs_Depth} left), if the cancer is well localized
the beam damages not only the sick tissue but also the healthy one. On the
contrary, electron therapy is particularly suited for skin and other superficial cancers,
because the dose of electrons (beta rays) practically vanishes beyond 1-2 cm, whereas
proton and more generally ion therapy is particularly suited for deeper cancers, because 
the dose of ions has its maximum (the {\it Bragg peak}) at a depth tunable up to 15-20
cm. In fact, in either case the collision cross-section with water molecules is strongly energy dependent, 
and the depth at which most  energy is deposited in human tissue can be fine-tuned.
Today 58 proton and 8 carbon ion therapy centers exist  (resp. 21, 3 in Europe); more are planned or under construction.
All have big size, high cost, high complexity; 
for instance, the CNAO hadron therapy center in Pavia uses
a 25m diameter synchrotron which has costed about 100 MEuro.
In fact, past and present-day acceleration techniques (cyclotrons, synchrotrons,...)
rely on the interaction of radio-frequency electromagnetic (EM) waves   with `few' charged 
particles (those one wishes to accelerate)   over long distances. The main reason of these structural limits is 
that electric fields cannot exceed the threshold of material breakdown (due to discharge sparks between electrodes) 
of $10\div 100$MeV/m, therefore 
accelerating an electron or a proton to 1 GeV  even by the  most powerful machines requires
a distance of 50-100 m. 
The search for alternative acceleration mechanisms is therefore of great importance.

\begin{figure}
\includegraphics[width=16cm]{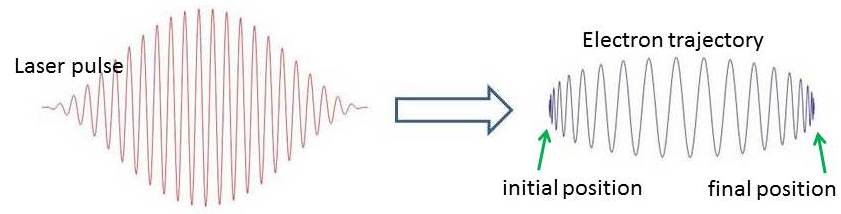}
\caption{Sample laser pulse and consequent motion in vacuum of an electron intially at rest.}
\label{vacmotion}       
\end{figure}

In vacuum, a coherent EM wave (laser pulse) reaching a charged particle at rest induces a motion 
composed of a transverse oscillation and a drift in the longitudinal direction $\vec{z}$ of propagation of the pulse, 
as depicted in fig. \ref{vacmotion}. This drift is caused by the  ponderomotive force 
$F\!_p\!:=\!\langle -e(\frac{\bv}c \times \bB)^z\rangle$
generated by the pulse; here   $\langle\: \rangle$ is 
 the average over a period of the laser carrier wave, $\bE,\bB$ are the electric and magnetic fields,
$\bv$ is the electron velocity, $c$ is the speed of light,  $\hat \bz$ is the direction of propagation
of the laser pulse; $F\!_p$  is positive (negative) while the modulating amplitude 
$\epsilon_s$ of the pulse  respectively grows (decreases).
During very intense laser pulses  the particle becomes  relativistic, but
under broad conditions its initial and final energies are practically equal, i.e. no net energy gain is possible; 
this is the socalled `Lawson-Woodward theorem' \cite{Woo47,Law79,Pal80,EsaSprKra95}.

One can try to evade the theorem by laser-matter, more precisely laser-plasma interactions
(by the way, very intense laser pulse locally ionize matter and convert it into a plasma).
An intense laser beam (alternatively, a beam of high energy protons/eletrons) travelling in a plasma  
causes large longitudinal charge density variations (lighter electrons are 
displaced with respect to heavier  ions) and thus a huge  longitudinal electric field $\bE$, 
due to the huge numbers of electrons and ions present. 
These variations arrange in a wake of waves ({\it plasma waves})  
traveling with phase velocity close to $c$: electrons are  several times  boosted forth and back, 
squeezed and unsqueezed, but are again left behind the laser beam with low speed.
This is similar to the fate of most water molecules in water waves. However,
if  some  foam  at  the crest of a water wave is a bit faster than the surrounding water, then 
it is accelerated `surfing' down the water wave slope (see fig. \ref{Dose_vs_Depth} right). Similarly,
 if some electrons are injected faster than their neighbours, they can be accelerated 
`surfing' down the plasma wake waves: this is the 
socalled {\it Wake-Field Acceleration}  (WFA) mechanism conceived by 
Tajima and Dawson \cite{TajDaw79}.
Such electrons are finally expelled out of the plasma sample behind 
the beam, in the \underline{same direction}. 
The WFA is especially effective in the  {\it bubble regime}, where
the `troughs' of the wake correspond to `ion bubbles' deprived of electrons:
it yields  nearly monochromatic and collimated electron bunches
of high energy. However, the onset of the bubble regime is not under full control yet.
Records established using laser pulses of wavelength $\lambda\!\sim\! 1\mu$m,
length $l\!\sim\! 10 \mu$m, energy $\E$, hitting helium jets of electron density 
$n_0=10^{17}\div\times 10^{19}\mbox{cm}^{-3}$ are:
\begin{itemize}

\item $200$ MeV electrons were obtained  in 2004 using  $\E\!\sim $1J laser pulses  \cite{ManEtAl04,GedEtAl04,FauEtAl04}.

\item $2\div 5$ GeV electrons were obtained in 2013-14  using   \
$\E\!\lesssim\! 150$J laser pulses (by a PetaWatt laser) \cite{WanEtAl13}.

\end{itemize}
 
\begin{figure}
\includegraphics[width=8.2cm]{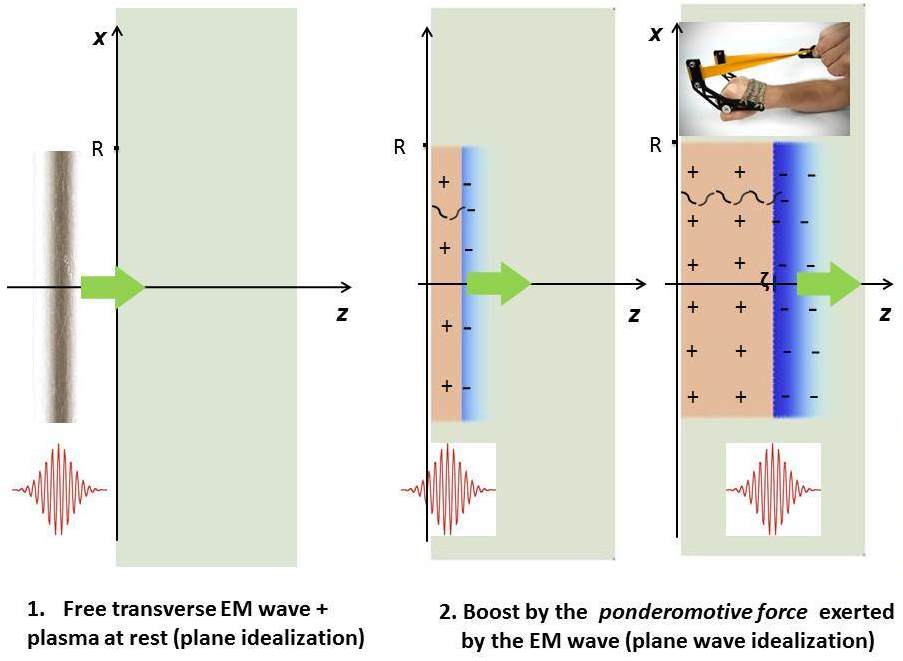} 
\includegraphics[width=8.2cm]{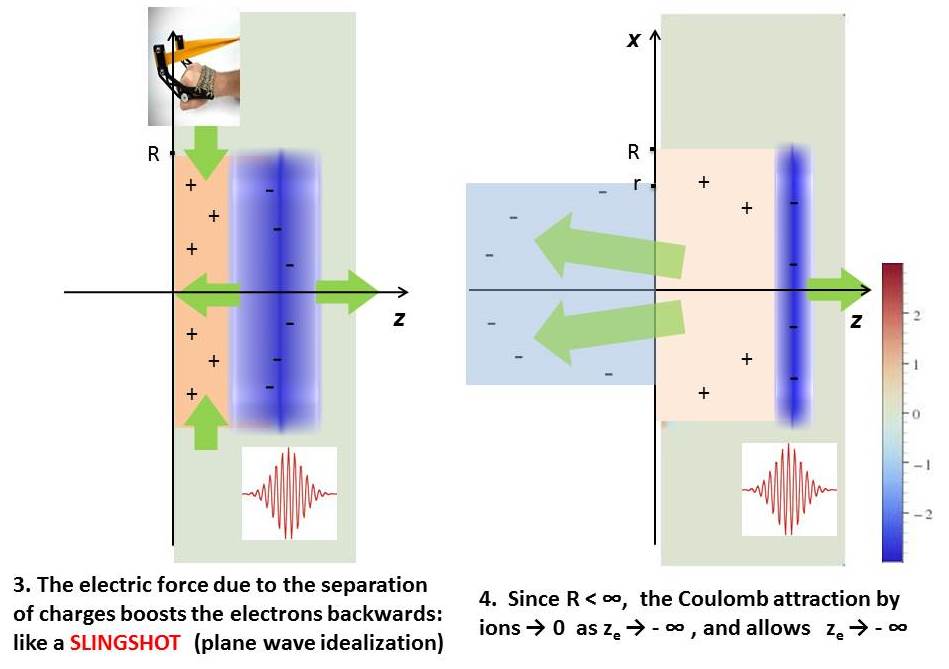}
\caption{Schematic stages of the slingshot effect}
\label{plasma-laser2}     
\end{figure}

We have recently suggested \cite{FioFedDeA14,FioDeN15}   the existence
of one more acceleration mechanism: the impact of a very short and intense  laser pulse 
in the form of a pancake normally onto 
the surface of a low-density  plasma may induce also the
acceleration and expulsion of electrons  \underline{backwards} \ ({\it slingshot effect}), see fig. \ref{plasma-laser2}.
A bunch of plasma electrons - in a thin layer just beyond the vacuum-plasma interface - 
first are displaced forward with respect to the ions  by the ponderomotive force 
generated by the pulse, then  are pulled back by the electric force $-eE^z$ due to this charge displacement.
Tuning the electron density $\widetilde{n_{0}}$ in the range where the
plasma oscillation period $T\!_{{\scriptscriptstyle H}}$\footnote{$T\!_{{\scriptscriptstyle H}}$ 
  grows with the oscillation amplitude $\zeta$, but goes to the nonrelativistic period 
$T\!_{{\scriptscriptstyle H}}^{{\scriptscriptstyle nr}}\!=\!\sqrt{\pi m/n_0e^2}$ as $\zeta\!\to\! 0$} 
is about twice the pulse duration $\tau$, 
we can make these electrons invert their motion when they are reached by the maximum of $\epsilon_s$, so that the negative part of $F\!_p$ adds to  $-eE^z$ in  accelerating them backwards; thus the total work 
$W\!=\!\int_0^\tau \! dt \, F\!_p \langle v^z\rangle $ done by the ponderomotive force is  maximal. 
The radius $R$  of the laser spot should be ``small'', for the pulse intensity 
- as well as the final energy of the expelled electrons escaping to $z\!\to\!-\infty$ - to be ``large'',  but 
not so small that lateral electrons obstruct them the way out backwards.
If $\tau\ll T\!_{{\scriptscriptstyle H}}$, then while the pulse is passing the electric force due to charge separation
can be neglected, and the motion of the electron 
is close to the one in vacuum (fig. \ref{vacmotion}); 
the backward acceleration takes place afterwards and is due only to   $-eE^z$, hence the final
energy is smaller. Whereas if \  $\tau\!\gg\! T\!_{{\scriptscriptstyle H}}$ - which was  the standard situation in laboratories 
until a couple of decades ago - then  
$F\!_p v^z$ oscillates many times about 0,  $W\!\simeq\!0$, and the slingshot effect is washed out.

Very short  $\tau$'s and huge nonlinearities
make   approximation schemes  based on  Fourier analysis and related methods
unconvenient. On the contrary, in  \cite{FioDeN15,Fio16} it is shown that
 in the relevant space-time region
a MFD description of the impact  is self-consistent, simple and predictive,
without need of a recourse to kinetic theory (i.e to a statistical description in phase space) taking
collisions into account, e.g. by BGK \cite{BhaGroKro54} equations or effective linear inheritance relations \cite{FioMaiRen14}. 
The set-up is as follows.
We regard the plasma as collisionless, with the ions at rest and 
 a fully relativistic  fluid of electrons; the system ``plasma + electromagnetic field"  fulfills
the Lorentz-Maxwell and the continuity Partial Differential Equations (PDE). 
For brevity, below we  refer to the electrons'
fluid element initially located at $\bX\!\equiv\!(X,Y,Z)$ as to the  ``$\bX$ electrons", and to the fluid elements with arbitrary $X,Y$ and specified $Z$  as the ``$Z$ electrons".
We denote: as \ $\bx_e(t,\bX)$  \ the position at time $t$
of the \ $\bX$ electrons, \ and for each fixed $t$ as
$\bX_e(t,\bx)$ the inverse of  $\bx_e(t,\bX)$  [$\bx\!\equiv\!(x,y,z)$]; as $m$ and
as $n,\bv,\Bp$ the electrons' mass and Eulerian
density, velocity, momentum.  \ $\bb\!:=\!\bv/c$,
$\bu\!:=\!\Bp/mc\!=\!\bb/\sqrt{1\!-\!\bb{}^2}$, 
$\gamma\!:=\!1/\sqrt{1\!-\!\bb{}^2}\!=\!\sqrt{1\!+\!\bu^2}$ \ are dimensionless.
Lagrangian fields carry a $\tilde{}$ and are related to Eulerian ones by the relation
$\tilde f(t,\bX)\!=\!f[t,\bx_e(t,\bX)]$.
We assume that the plasma is initially 
neutral, unmagnetized and at rest with electron (and proton) density $\widetilde{n_{0}}(Z)$
depending only on $Z$ and
equal to zero  in the region  \ $Z\!<\! 0$. 
We schematize the laser pulse  as a free transverse EM plane travelling-wave
 multiplied by a cylindrically symmetric ``cutoff'' function, e.g.
\be
\bE^{{\scriptscriptstyle\perp}}\!(t,\!\bx)=\Be\!^{{\scriptscriptstyle\perp}}\!(ct\!-\!z)\,
\theta(R\!-\!\rho),\qquad  \bB^{{\scriptscriptstyle\perp}}=
{\hat{\bf z}}\!\times\!\bE\!^{{\scriptscriptstyle\perp}}                                                  \label{pump}
\ee
where  $\rho\!:=\sqrt{\!x^2\!+\!y^2}\!\le\! R$, 
$\theta$ is the Heaviside step function, and
the  `pump'  $\Be^{{\scriptscriptstyle\perp}}(\xi)$ \ vanishes
outside some finite interval \ $0\!<\!\xi\!<\!l$.
Then, to simplify the problem:
\begin{enumerate}

\item We study the $R\!=\!\infty$ (i.e. {\it plane-symmetric}) version first,
carefully choosing  unknowns and  independent variables  (sect. \ref{Planewavessetup}).
For small $\widetilde{n_{0}}(Z)$ and short times 
 we can reduce  the PDE's to a  collection of decoupled {\it systems of two 1st order 
nonlinear ODE in  Hamiltonian form}, which we  solve numerically.

\item  We determine (sect. \ref{3D-effects}): 
$R\!<\!\infty$ and a suitable positive $r\!\le\!R$ so that the plane version   gives small errors at least
for $\bX$ electrons  with $Z\!\gtrsim\! 0$ and $\sqrt{X^2\!+\!Y^2}\!\le\!r$ until their expulsions;
sample final energies, spectra of the expelled electrons. 
For definiteness, we consider $\widetilde{n_{0}}(Z)$ of the kind of fig. \ref{graphsb}.

\end{enumerate}

\section{Model and predictions}
\subsection{Plane wave idealization}
\label{Planewavessetup}

\noindent
Here is our plane wave Ansatz: \ $A^\mu$ (the EM potential), $\bu,n\!-\!\widetilde{n_{0}}(z)$ \ depend only on  $z,t$
and vanish if \  $ct\!\le\! z$; \ 
$\Delta \bx_e\!:=\! \bx_e\!-\!\bX$ \  depends only on $Z,t$ and vanishes 
 if \  $ct\!\le\! Z$. Then:   $\bB\!=\!\bB^{{\scriptscriptstyle\perp}}\!=\!\hat\bz\!\wedge\!\partial_z\bA\!^{{\scriptscriptstyle\perp}}\!,$ \    
 $c\bE^{{\scriptscriptstyle\perp}}\!\!=\!-\partial_t\bA\!^{{\scriptscriptstyle\perp}}\!$; \  the transverse component of the Lorentz equation \ $d {\bf p}/dt =
-e(\bE +\frac{\bv{{}}}c \wedge \bB)$ \ and the initial condition ${\bf p}\!\equiv\!0$ imply
$\bu^{{\scriptscriptstyle\perp}}\!= \!e\bA\!^{{\scriptscriptstyle\perp}}\!/m{}c^2$; by the continuity equation
the Eulerian electron density $n_e$ and the initial one $\widetilde{n_{0}}$
are related through \ $n_e(t,\!z)= \widetilde{n_{0}}\!\left[Z_e(t,\!z)\right]\, \partial  Z_e(t,\!z)/\partial z$;  
\ by the Maxwell
equations $E^z$ is determined by the longitudinal motion and $\widetilde{n_{0}}$ through
\be
E^z(t,\! z)\!=\!4\pi e \!\left\{
\widetilde{N}(z)\!-\! \widetilde{N}[Z_e (t,\! z)] \right\}, \qquad 
\widetilde{N}(Z)\!:=\!\!\!\int^Z_0\!\!\!\!\!\! d\eta\,\widetilde{n_{0}}(\eta);
 \label{elFL}{}
\ee
the positive (resp. negative) term at the right-hand side is  due to the ions (electrons).
For sufficiently small densities and short times the laser pulse is 
not significantly affected by the interaction with the plasma 
(the validity of this approximation is checked a posteriori \cite{FioDeN15}),
and we can identify  \ $\bA\!^{{\scriptscriptstyle\perp}}(t,z)\!=\!\Ba(\xi)$,  
$\bu\!^{{\scriptscriptstyle\perp}}(t,z)\!=\!e\Ba(\xi)/m{}c^2\!=:\!\hat\bu\!^{{\scriptscriptstyle\perp}}(\xi)$ \ where
 $\xi\!:=\! ct\!-\!z$, \ and $\Ba(\xi)\!:=\!-\int_0^\xi\!ds\Be\!^{{\scriptscriptstyle\perp}}\!(s)$ is the
transverse vector potential of the `pump' free laser pulse.
The remaining unknowns are $u^z(t,z)$ and $z_e(t,Z)$.
In the equations of motion of the $Z$-electrons  (Lagrangain description)
$z_e(t,Z)$ appears everywhere in place of $z$, e.g.
the force associated to (\ref{elFL}) is
$\widetilde{F}_e^{{\scriptscriptstyle z}}\!(t,\!Z)\!:=\!4\pi e^2 \!\left\{\!\widetilde{N}\!(Z)
\!-\! \widetilde{N}\![z_e(t,Z)]\!\right\}$.  This is conservative,
since it depends on $t$ only through $z_e(t,Z)$. 
As no particle can reach the speed of light,  the map  $t\!\mapsto\! \xi\!=\!\tilde \xi(t,\!Z)\!:=\! ct\!-\!z_e(t,\!Z)$  is strictly increasing, and we can  
use  {\it $(\xi,Z)$  instead of}
$(t,\!Z)$ as independent variables, so that the argument $\xi$ of $\hat\bu\!^{{\scriptscriptstyle\perp}}$ 
is; we shall denote  the dependence of a field on $(\xi,Z)$ by a caret.  
It is also convenient to use the \ `electron $s$-factor'  $\hat s\!:=\!\hat \gamma\!-\! \hat u^z$ \
instead of \ $\hat u^z$ \ as an unknown, because 
$\hat \gamma,\hat \bu,\hat \bb$ are {\it rational}  functions (no square roots!) of $\hat \bu^{{\scriptscriptstyle\perp}},\hat s$,
\be
\hat \gamma\!=\!\frac {1\!+\!\hat \bu^{{\scriptscriptstyle\perp}}{}^2\!\!+\!\hat s^2}{2\hat s}, 
\qquad \hat u^z\!=\!\frac {1\!+\!\hat \bu^{{\scriptscriptstyle\perp}}{}^2\!\!-\!\hat s^2}{2\hat s},
 \qquad \hat \bb\!=\! \frac{\hat \bu}{\hat \gamma} \label{u_es_e}
\ee
(these relations hold also with the caret replaced by a tilde or nothing),
and - as we will show - $\hat s$ is {\it insensitive} to rapid oscillations of $\Ba$. \ 
The definitions \ $\tilde\bb\!:=\!\tilde\bv/c\!=\!\partial \bx_e/c\partial t$ and $\tilde s\!:=\!\tilde \gamma\!-\! \tilde u^z\!=\!\sqrt{1\!+\!\tilde \bu^2}\!-\! \tilde u^z$, together with the Lorentz equation (in Lagrangian formulation) $\partial \tilde\bu/\partial t =
-e(\tilde\bE \!+\!\tilde\bb{{}} \wedge \tilde\bB)/mc$, lead to
\bea
\frac 1c \frac{\partial (\bx_e\!-\!\bX)}{\partial t}=\tilde \bb\!\stackrel{(\ref{u_es_e})}{=}\!\frac {\tilde\bu}{\tilde \gamma},
 \qquad
\tilde \gamma\frac{\partial \tilde s}{\partial t} =-\frac{\tilde F_e^{{\scriptscriptstyle z}}}{mc} \tilde s\!+\! \tilde g
=\frac{4\pi e^2}{mc} \!\left\{\!\widetilde{N}[z_e(t,\!Z)]
\!-\! \widetilde{N}(Z)\!\right\}\!\tilde s\!+\! \tilde g, \nonumber
\eea
where $g\!:=\!(\partial_t\!+\!c\partial_z)\bu^{{\scriptscriptstyle\perp}2}$. Since $g\!\equiv\!0$ in our approximation 
$\bu\!^{{\scriptscriptstyle\perp}}\!(t,z)\!=\!\hat\bu\!^{{\scriptscriptstyle\perp}}(ct\!-\!z)$ and $c\hat s\partial/\partial\xi\!=\!\tilde\gamma\partial/\partial t$, then, 
switching to the independent variables $\xi,Z$ we find  
\bea
&& (\hat\bx_e^{{\scriptscriptstyle\perp}}\!-\!\bX^{{\scriptscriptstyle\perp}})'=
\hat\bu^{{\scriptscriptstyle\perp}}/\hat s,\qquad  
\hat\bx_e^{{\scriptscriptstyle\perp}}(0,\!Z)\!-\!\bX^{{\scriptscriptstyle\perp}}\!=\!0,         \label{utileperp} \\[8pt]
&&\hat \Delta'=\displaystyle\frac {1\!+\!v}{2\hat s^2}\!-\!\frac 12,
\qquad \hat s'=\frac{4\pi e^2}{mc^2}\left\{\!
\widetilde{N}[\hat\Delta\!+\!Z] \!-\! \widetilde{N}(Z)\!\right\}\quad \label{heq1}  \\[8pt]
&&\hat \Delta(0,\!Z)\!=\!0,  \qquad\qquad
 \hat s(0,\!Z)\!=\! 1. \qquad\qquad\qquad\qquad  \label{heq2}
\eea
Here  \ $\hat \Delta(\xi,Z)\!:=\! \hat z_e\!(\!t\!,\! Z\!)\!-\!\! Z$ is the electrons' longitudinal displacement
with respect to the initial equilibrium position $Z$, \ $f'\!:=\!\partial f/\partial \xi$, \ $v(\xi )\!:=\!\hat\bu\!^{{\scriptscriptstyle\perp}2}(\xi)$.

The PDE to be solved are reduced to the
collection (\ref{heq1}-\ref{heq2}) of systems  (parametrized by $Z$) of  first order ODE's in the unknowns
$\hat \Delta(\xi,Z)$,  $\hat s(\xi,Z)$. In fact we now show how determine  all unknowns once  (\ref{heq1}-\ref{heq2}) is solved.
Let
\bea
\ba{ll}
\hat \bY^{{\scriptscriptstyle\perp}}\!(\xi,\!Z)\!:=\!\!\displaystyle\int^\xi_0\!\!\!\!\! d\xi' \frac{\hat \bu^{{\scriptscriptstyle\perp}}\!(\xi')}{\hat  s(\xi'\!,\!Z)},\qquad &\hat Y^z\!(\xi,\!Z)\!:=\!\!\displaystyle\int^\xi_0\!\!\!\!\! d\xi' \frac{\hat u^z\!(\xi'\!,\!Z)}{\hat  s(\xi'\!,\!Z)},\\[8pt] 
\hat \Xi(\xi,\!Z)\!:=\!\!\displaystyle\int^\xi_0\!\!\!\!\!  d\xi' \frac{\hat \gamma(\xi'\!,\!Z)}{\hat s(\xi'\!,\!Z)},
\qquad &  c\hat t(\xi,\!Z)\!:=\!\xi\!+\!\hat z_e(\xi,Z); 
\ea\quad                      \label{defYXi}               
\eea
by (\ref{u_es_e}), (\ref{heq1}) it is also $\hat Y^z\!\equiv\!\hat \Delta$, $\hat \Xi(\xi,\!Z)\!=\!\xi  \!+\! \hat \Delta(\xi,\!Z)$,
$ c\hat t(\xi,\!Z)\!=\!Z \!+\! \hat \Xi(\xi,\!Z)$. \ One immediately checks that 
$\hat \Xi(\xi,\!Z)$ is strictly increasing (hence invertible) with respect to $\xi$ for all fixed $Z$, 
eq. (\ref{utileperp}) is solved by \ $\hat\bx_e^{{\scriptscriptstyle\perp}}(\xi,\!Z)\!-\!\bX^{{\scriptscriptstyle\perp}}\!=
\!\hat \bY^{{\scriptscriptstyle\perp}}\!(\xi,\!Z)$, and
$ \hat t(\xi,\!Z)$ is the inverse of $\tilde\xi(t,Z)\!:=\!ct\!-\!z_e(t,Z)$. Note that both
\be
\hat\bx_e(\xi,\!\bX)\!=\!\bX\!+\!\hat \bY\!(\xi,\!Z), \qquad  c\hat t(\xi,\!Z)\!=\!\xi\!+\!\hat z_e(\xi,Z)
\!=\!Z \!+\! \hat \Xi(\xi,\!Z),
\ee
can be also obtained  solving for $t,\bx$ vs. $\xi,\bX$ the system of functional equations  
\be
\xi=ct\!-\!z,  \qquad  
\hat \Xi(ct\!-\!z,\!Z)\!=\!ct\!-\!Z,  \qquad  \hat \bx-\bX= \hat \bY(ct\!-\!z,\!Z)                         \label{functeq}
\ee
[the second is actually equivalent to the $z$-component of the third]. In general one can solve 
(\ref{functeq}) in four ways  for one out of $\{t,\xi\}$ and one out of
$\{\bx,\bX\}$ as functions of the remaining two  variables;
thus  one finds the original unknowns
\bea
 \tilde\xi(t,Z)\!=\!
\hat \Xi^{{{\scriptscriptstyle -1}}}(ct\!-\!Z,\!Z),\qquad
 \bx_e(t,\!\bX\! )\!=\!\bX\!+\!\hat \bY\!\!\left[\hat \Xi^{{{\scriptscriptstyle -1}}}\!(ct\!-\!Z,\!Z),\!Z\!\right]
   \qquad        \label{sol}
\eea
and other useful relations obtained by derivations. 
In particular, $\partial_{{\scriptscriptstyle Z}}\hat z_e\!\equiv\!1\!+\!\partial_{{\scriptscriptstyle Z}}\hat\Delta\!>\!0$ is 
a necessary  and sufficient condition for
the invertibility  (at fixed $t$) of the maps \ $z_e\!:\!Z\!\mapsto\! z$,  $\bx_e\!:\!\bX\!\mapsto\!  \bx$, justifying the 
MFD description adopted so far.  
By replacement we obtain the other unknowns $\bu,\gamma,\bb,n_e$, e.g.
\bea
&&\tilde\bu(t,\!Z )\!=\!\hat\bu\!\!\left[\hat \Xi^{{{\scriptscriptstyle -1}}}\!(ct\!-\!Z,\!Z),\!Z\!\right]\!, \quad 
\bu(t,\!z )\!=\!\hat\bu\!\left[ct\!-\!z,\!Z_e(t,\!z)\!\right]\!, \\        
 &&  n_e(t,z)\!=\! 
\widetilde{n_0}\!\left[Z_e(t,\!z)\right]\left.\frac{\hat\gamma}{\hat s\,\partial_Z\hat z_e}\right\vert_{(\xi,Z)=
\big(ct\!-z,Z_e(t,z)\big)}\!\!.     \qquad    \label{expln_e}
\eea
We have thus shown that solving (\ref{heq1}-\ref{heq2}) (e.g. numerically) and inverting the functions
$\xi\!\mapsto\!\hat\Xi(\xi,Z)$, $Z\!\mapsto\! z_e(t,Z)$ all unknowns can be determined explicitly. 

An alternative derivation of (\ref{utileperp}-\ref{heq2}) 
 from the least action principle is given in \cite{Fio16'}.

\begin{figure}
\includegraphics[width=8.2cm]{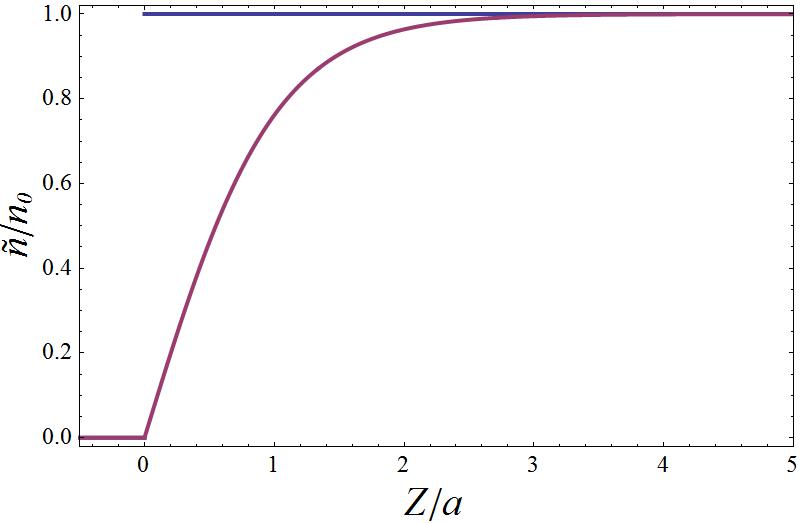} \hfill
\includegraphics[width=8.2cm]{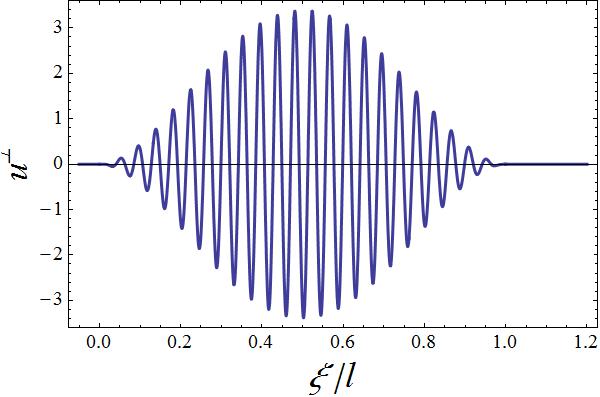} \\
\includegraphics[width=8.2cm]{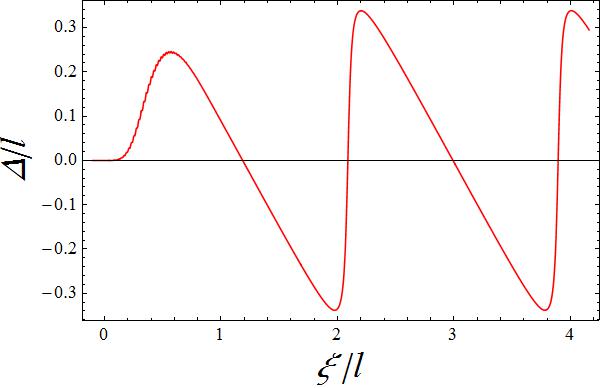} 
\hfill\includegraphics[width=8.2cm]{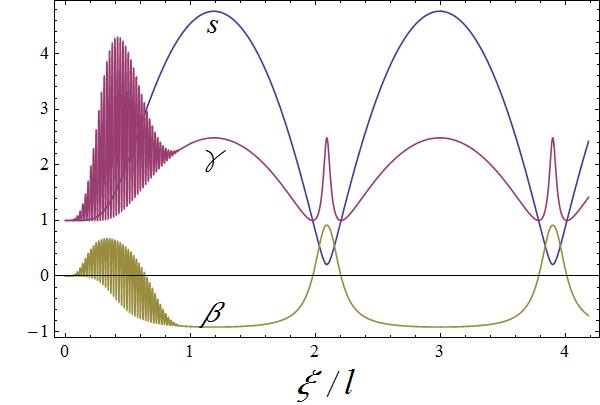} \\
\includegraphics[width=6.75cm]{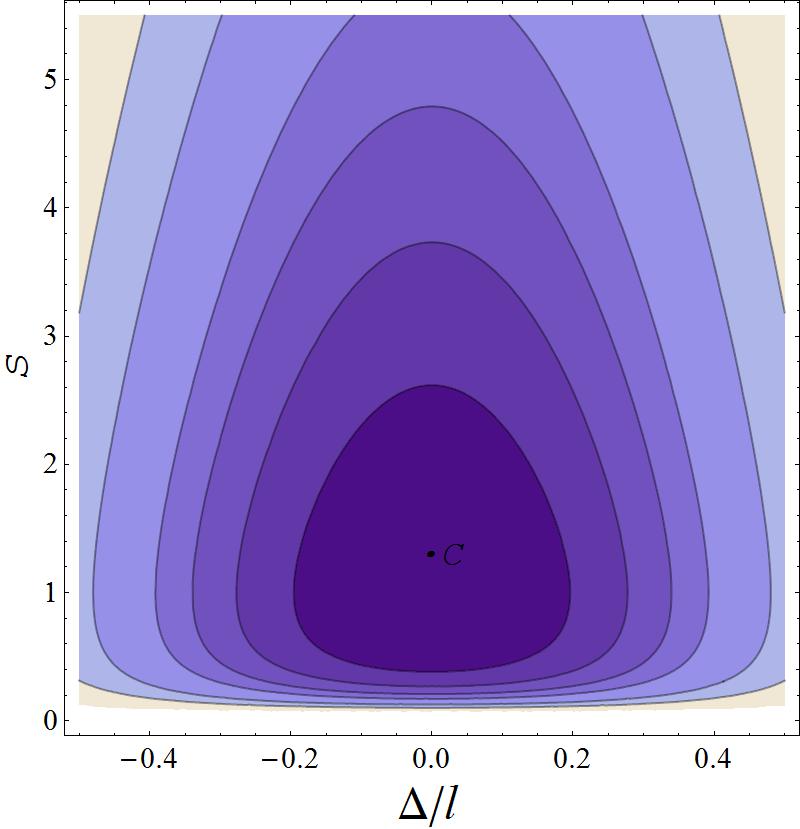}\hskip.2cm
\includegraphics[width=10.5cm]{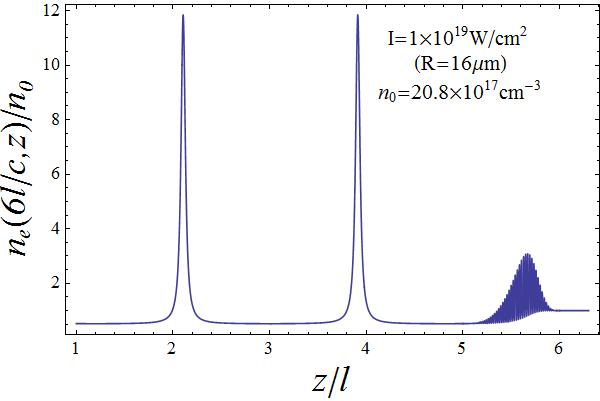}
\caption{Up-left: the normalized $\widetilde{n_{0}}$'s adopted here: step-shaped  (blue) or
continuous  $\widetilde{n_{0}}(Z)\!=\!n_0\,\theta(Z)\tanh(Z/a)$,   $a\!=\!20\mu$m (purple); they 
respectively model the initial electron  densities at the vacuum interfaces of an aerogel and of a gas jet
(just outside the nozzle). Up-right: normalized pump amplitude \ $\hat\bu^{{\scriptscriptstyle\perp}}\!=\!e\Ba/mc^2$  
of a pulse as in table \ref{tab1g}, $R\!=\!16\mu$m ($\hat\bu^{{\scriptscriptstyle\perp}}(\xi)\!=\!0$  outside 
$0\!<\!\xi \!<\!l\!=\!18.75\mu$m). 
Center:  corresponding solution of (\ref{e1}-\ref{e2})  for   $Ml^2\!=\!26$
(i.e. $n_0\!=\! 20.8 \!\times\! 10^{17}$cm$^{-3}$; down-right: corresponding
electron density for $z\!\gtrsim\!20\mu$m about 600 fs after the impact of the laser on the plasma. Down-left:  
phase space paths of  (\ref{e1}-\ref{e2}) with $v\!\equiv\!0$,  $Ml^2\!=\!26$ and growing values of the total energy.}
\label{graphsb}
\end{figure}

Even though $\Be^{{\scriptscriptstyle\perp}},\hat\bu^{{\scriptscriptstyle \perp}}$ oscillate fast with $\xi$, since $v\!\ge\! 0$ 
integrating (\ref{heq1}) makes relative oscillations of $\hat\Delta$ much smaller  than those of $v$ and
 those of $\hat s$ much smaller than the former; hence, as anticipated, $\hat s$ is practically
smooth, see e.g. fig. \ref{graphsb}. In vacuum ($\widetilde{n_{0}}\!\equiv\!0$)  it is even
$\hat s\!\!\equiv\!\!1$, and the equations are solved in closed form \cite{Fio14JPA,Fio14}. 
Note also that the right-hand side of (\ref{heq1})$_2$ is an increasing function of $\hat\Delta$, because so is $\widetilde{N}(Z)$.
Therefore, as  $v(\xi )$ \ is zero for $\xi\!\le\!0$ and positive  for $\xi\!>\!  0$,  then so are also
$\hat\Delta(\xi,\!Z)$ and $\hat s(\xi,\!Z)\!-\!1$. \  Both keep increasing 
until $\hat\Delta$ reaches a positive maximum $\hat\Delta(\bar\xi,Z)$ 
at the $\xi\!=\!\bar\xi(Z)\!>\!0$ such that $\hat s^2(\bar\xi,\!Z)\!=\!1\!+\!v(\bar\xi)$   ($\bar\xi\!<\!l$ if $v(l)\!=\!0$),
see  fig. \ref{graphsb}.
The time of maximal penetration of the $Z$ electrons is thus \ $\bar t(Z)\!=\![Z \!+\! \hat \Xi(\bar\xi,\!Z)]/c$.

Eq.s (\ref{heq1}) can be written also in the form \cite{Fio16} of  {\it Hamilton equations}  
\ $q'=\partial H/\partial p$, $p'=-\partial H/\partial q$ \ in 1 degree of freedom:
  $\xi,-\hat\Delta, \hat s$  play the role of $t,q,p$, and the Hamiltonian reads
\bea
\ba{l}
H\!(\Delta,\!s,\!\xi;\!Z):= \gamma(s,\xi)+ \U(\Delta;\!Z),\quad\gamma(s,\xi)\!:=\! 
[s^2\!\!+\!1\!+\!v(\xi)]/2s ,\\[8pt]
 \U(\Delta;\!Z)\!:=\!\frac{4\pi e^2}{mc^2}\!\left[
\widetilde{{\cal N}}\!(Z \!+\!\Delta) \!-\!\widetilde{{\cal N}}\!(Z)\!-\! \widetilde{N}\!(Z)\Delta\right]\!,\quad
U(z_e;\!Z)\!:=\!\U(\Delta\!+\!Z;\!Z),\\[8pt]
 \widetilde{{\cal N}}(Z):=
\int^Z_0\!\!\!dZ'\,\widetilde{N}(Z')\!=\!\int^{Z}_0\!\!\!dZ'\, \widetilde{n_0}(Z')\, (Z\!-\!Z').
\ea                                \label{hamiltonian}
\eea
Defining $\U$  we have fixed the free additive constant so that  $\U(0,\!Z)\!\equiv\! 0$ for each $Z$,
\ $H\!-\!\sqrt{1\!+\!v}$ \ is positive definite.

\medskip
If in particular $\widetilde{n_{0}}(\!Z\!)\!=\!n_0\theta(\!Z\!)$ (step-shaped initial density), then by (\ref{elFL})
 the  longitudinal electric force acting on the $Z$-electrons is
\be
\hat F_e^{{\scriptscriptstyle z}}(\xi,Z)\!=\! \left\{\!\!\ba{ll}
- 4\pi n_0 e^2\!  \hat\Delta(\xi,Z)  \mbox{= elastic force}\quad &\mbox{if }\hat z_e\!=\!Z\!+\!\hat\Delta \!>\!0,\\[6pt]
 \: 4\pi n_0 e^2 Z\: \mbox{= constant force}\quad &\mbox{if }\hat z_e\!=\!Z\!+\!\hat\Delta\!\le\!0;
\ea\right.
\ee
hence   {\it as long as $z_e\ge 0$} each $Z$-layer of electrons  is an independent copy of the {\it same} 
relativistic harmonic oscillator,  (\ref{heq1}-\ref{heq2}) are $Z$-independent and 
(setting $ M \!:=\!4\pi n_0e^2\!/mc^2$) reduce 
to a {\it single} system of two  first order ODE's
\bea
&& \Delta'=\displaystyle\frac {1\!+\!v}{2 s^2}\!-\!\frac 12,\qquad\qquad
 s'=M\Delta,\label{e1}\\[6pt]
&&   \Delta(0)\!=\!0, \qquad\qquad\qquad\:\:   s(0)\!=\! 1;\label{e2}
\eea
correspondingly, the inverse function $Z_e(t,\!z)$ has the closed form  
\be
 Z_e(t,z)=ct\!\!-\!\Xi(ct\!-\!z)=z\!\!-\!\Delta(ct\!-\!z).                                               \label{sol'}
\ee 
In fig. \ref{graphsb} we plot a typical pump and the corresponding solution of (\ref{e1}-\ref{e2}); for
$\xi\!\ge\!l$ $v(\xi)\!=\!v(l)\!\equiv$const$\simeq\!0$, the equations become autonomous,
 all paths in phase space become cycles  around the center  
$C\!:=\!(\Delta,s)\!=\!(0,\sqrt{1\!+\!v(l)})$, and the solutions periodic of period $cT\!_{\scriptscriptstyle H}$;
hence the final result of the pulse interaction is to move the electrons from the center to a cycle of higher energy.

\subsection{Finite $R$ corrections and experimental predictions}
\label{3D-effects}

\begin{figure}
\includegraphics[width=7.2cm]{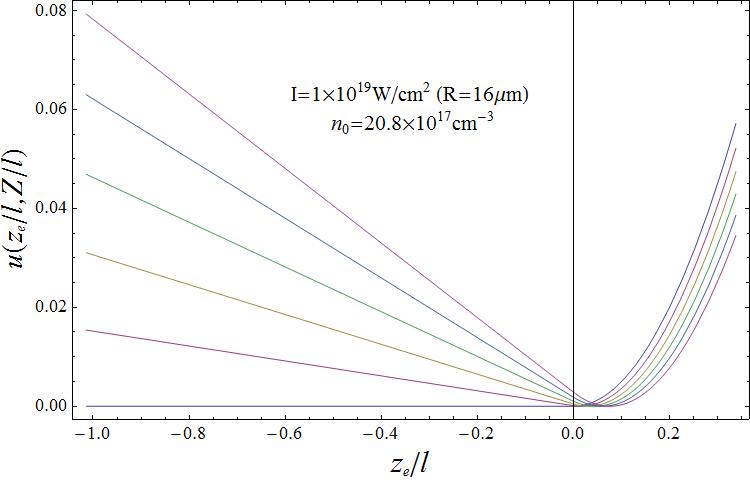} \hfill
 \includegraphics[width=8.6cm]{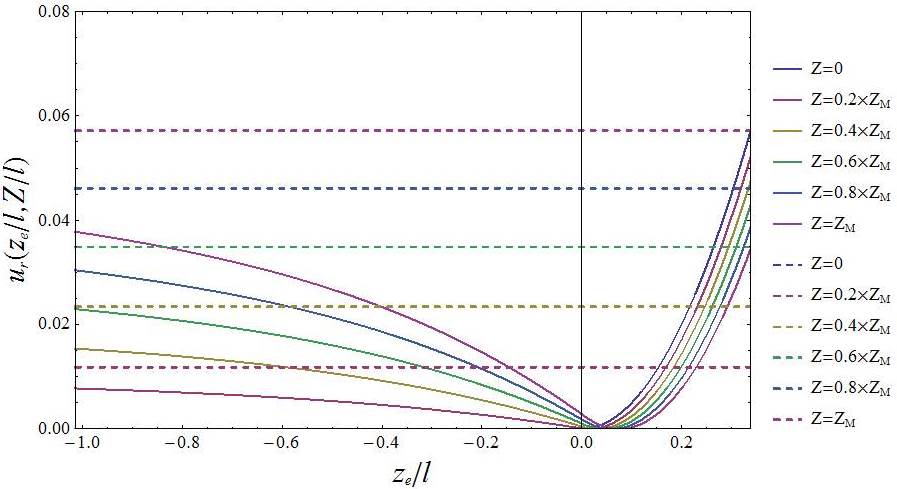}
\caption{Rescaled longitudinal electric potential energies
in the idealized plane wave  (left) and in the $R\!=\!16\mu$m (right)  case,
plotted as functions of $z_e$ for
$Z/\ZM=0,.2,.4,.6,.8,1$; the horizontal dashed lines are the left asymptotes of $u_r$ for
the same values of $Z/\ZM$.}
\label{Upotentials}
\end{figure}
\begin{figure}
\includegraphics[width=16cm]{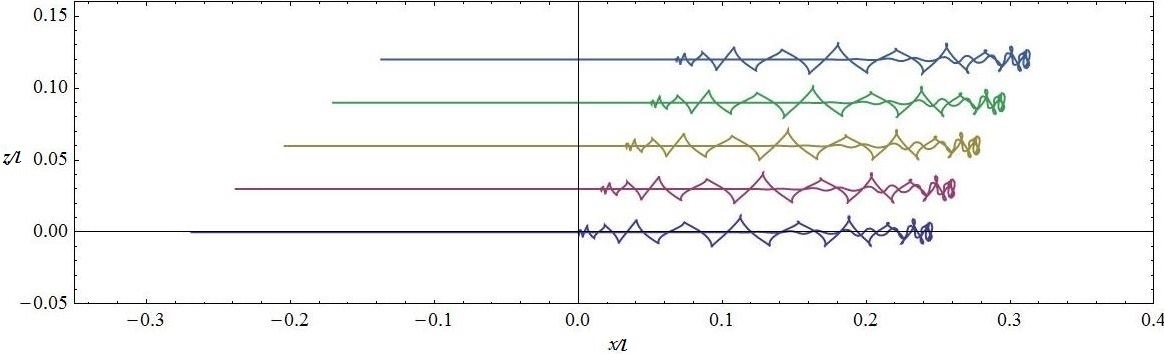} \\
\includegraphics[width=8.2cm]{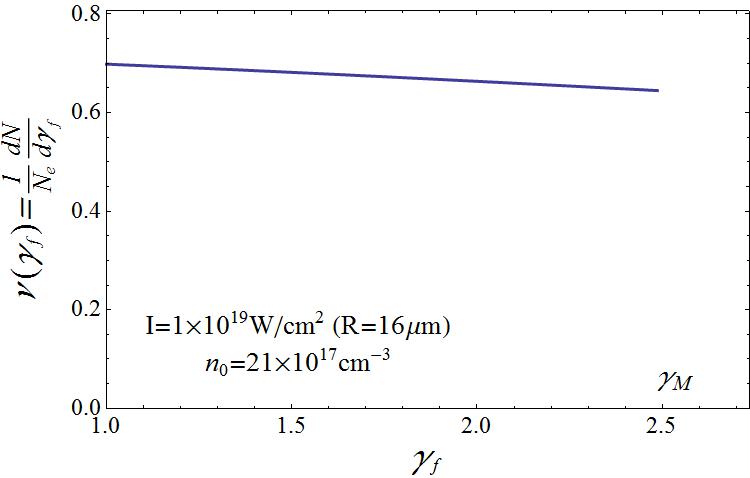} \hfill
\includegraphics[width=8.2cm]{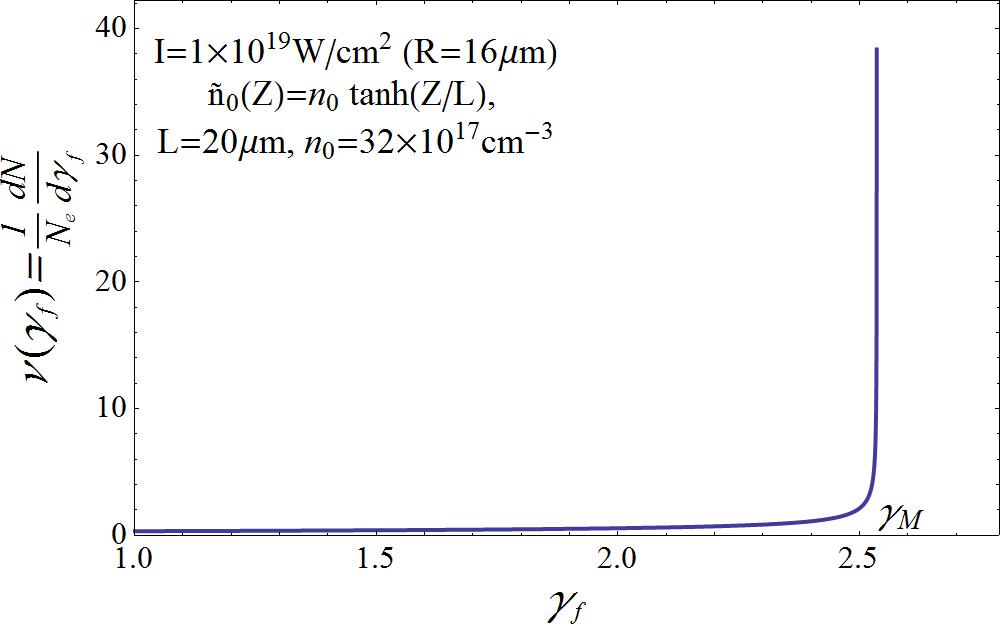} 
\caption{Up: trajectories  (after 150 fs) of electrons initially located at 
\  $Z/\ZM=0,  0.25 , 0.5 ,  0.75 ,1$  under the same conditions as in fig. \ref{graphsb}. Down:
Spectra of the  expelled electrons for average pulse intensitiy $I\!=$10$^{19}$ W/cm$^2$ and 
some step-shaped (left) or continuous (right) initial density $\widetilde{n_{0}}$.}
\label{Traj2}
\end{figure}

Since $R\!<\!\infty$  the  potential energies (parametrized by $Z\!>\!0$) $U(z_e,Z)$ associated to (\ref{elFL}) - due to
charge separation -  are  inaccurate as $z_e \!\to\! -\infty$. Hence we replace
 \ $U\!\mapsto\!\rU$ \ in the equations of motion, where $\rU$ is a suitable effective potential
differing from $U$ for $z_e \!<\! 0$; this  allows \   
$z_e(t,\!Z)\!\stackrel{t\to\infty}{\longrightarrow}\! -\infty$ \  (backward
 escape) for electrons in a suitable surface 
layer  $0\!\le\!Z\!\le\!Z_{{\scriptscriptstyle M}}$.
If e.g.  $\widetilde{n_{0}}(\!Z\!)\!=\!n_0\theta(\!Z\!)$ then $U,\rU$ (plot in fig. \ref{Upotentials})  
are given by
$U(z_e,Z)\!=\!2\pi n_0e^2[ \theta(z_e)z_e^2\!-\!2z_eZ\!+\!Z^2]$  and 
$$
\ba{l}
\rU(z_e,\!Z) \!=\!  \pi n_0e^2 \!\left[(z_e\!-\!2Z)\sqrt{\! (z_e\!-\!2Z)^2\!+\!R^2}\!-\!4Zz_e\!+\!
R^2\sinh^{-1}\!\!\frac{z_e\!-\!2Z}R  \right. \\[6pt]
\qquad\quad  \left.  -\!z_e\sqrt{\! z_e^2\!+\!R^2}\!-\!R^2\sinh^{-1}\!\!\frac{z_e}R\!+\! 2Z^2  \!+\!2Z\sqrt{\! 4Z^2\!+\!R^2} 
\!+\!R^2\sinh^{-1}\!\!\frac{2Z}R\right].
\ea
$$
Solving the equations the map \  $\bX\!\mapsto\! \bx_e(t,\bX)$ \ turns out to be  one-to-one
for all $t$
and either sufficiently small or sufficiently large $Z$, showing the {\it self-consistency} of this MFD treatment.
This was not granted by the equations alone: the invertibility of the map $\bX\!\mapsto\! \bx_e(t,\bX)$ 
fails with a large class of initial conditions \cite{Daw59}\footnote{With the 
initial conditions (\ref{heq2}) and a non-vanishing $v$ as considered here the invertibility of the map  $\bX\!\mapsto\! \bx_e(t,\bX)$
breaks also in an intermediate $Z$-range ($\ZM\!\le\! Z\!\le\!\ZM'$) for 
$t\!\gtrsim\!T\!_{{\scriptscriptstyle H}}$.}. For instance, in the
 non-relativistic limit (\ref{e1}) with $v\!\equiv\!0$ is equivalent to \ $\hat\Delta''\!=\!-M\hat\Delta$, \ which with the conditions
 $\hat\Delta(0;\!Z)\!\equiv\!0$,
$\hat\Delta'(0;\!Z)\!\equiv\!a Z$ is solved by
$$
z_e(\xi,\!Z)\!-\!Z=\Delta(\xi;\!Z)=aZ\sin(\sqrt{\!M} \xi), \quad\quad\Rightarrow \quad  
\frac{\partial \hat z_e}{\partial Z}=1\!+\!a\sin(\sqrt{\!M}\xi);
$$
hence \ $\partial \hat z_e/\partial Z \!=\!1$ at $\xi\!=\!0$,\ but  if $|a|\!>\!1$  
then $\partial \hat z_e/\partial Z\!<\!0$ at  sufficiently large $\xi$ (or times).
Sample trajectories of small $Z$ electrons are shown in fig.
\ref{Traj2}.
The interplay of the ponderomotive, electric forces
yield the longitudinal  forward and backward drifts at the basis of the slingshot effect.
On the contrary, transverse oscillations due to $\bE^{{\scriptscriptstyle \perp}}$ average to zero and yield  vanishing final transverse drift and momentum,
if - as usual - the pump (\ref{pump}) (here polarized in the $x$-direction) 
\be
\Be\!^{{\scriptscriptstyle\perp}}\!(\xi)
\!=\!{\hat\bx}\epsilon_s(\xi)\cos k\xi\qquad
\mbox{with }\: |\epsilon_s'|\!\ll\! |k\epsilon_s|
\label{modulation}
\ee
has a slow modulation $\epsilon_s$  in the support $0\!<\!\xi\!<\! l$; this
implies \   $p^{{\scriptscriptstyle \perp}}(\xi)\!\simeq\!
\epsilon_s(\xi) \left|\sin(k\xi) e/kc\right|\!=\!
0$ \ for $\xi\!\ge\! l$, and hence a good collimation of the expelled electrons.  
If the plasma is created by the impact on a supersonic gas jet (e.g. helium)  of the pulse itself, then
 $l\!<\!\infty$ is the length of the interval where the intensity is sufficient to 
ionize the gas.
The EM energy $\E$ carried by a pulse (\ref{pump}), (\ref{modulation})  is 
\be
\E\!\simeq\!\frac{R^2}{8}\!\! \int_0^l\!\!\!\!d\xi\,\epsilon_s^2(\xi).
\label{pulseEn}
\ee
$\E$ depends on the laser; 
reducing $R$ (focalization)  increases the intensity $I$, the electron 
penetration $\zeta$ and the slingshot force. 
But we need to tune $R$  so that  $\rU$ be justified, i.e. the ``information about the finite $R$" 
(contained in the retarded fields generated by charge separation) reach the $\vec{z}$-axis around the expulsion time $t_e$
(neither much earlier, nor much later). Moreover,  $R$ must be large enough for
 the Forward Boosted Electrons (FBE)  in an inner cylinder $\rho\!\le\! r\!\le\! R$ to be expelled before
Lateral Electrons (LE), initially located outside the surface of the 
 hole $\RC$ created by the pulse and attracted towards the $\vec{z}$-axis, obstruct their way out. 
These conditions amount to  \cite{FioDeN15,{Fio16a}}
\be
 \frac{[t_e \!-\!\bar t]c}R \sim 1,   \qquad
r := R-\frac{\zeta(t_e\!-\!l/c)}{2(t_e\!-\!\bar t)}\: \theta(ct_e\!-\!l ) >0,                                   \label{req}
\ee
which can be fulfilled also with a rather small $R$,  by the  delay inherent to the retarded potential itself
and the fact that contributions by ions and FBE sum up on their surface  
of separation, while they partially cancel on LE.

We report in table \ref{tab1g}  and fig.  \ref{Traj2} sample results of extensive numerical simulations
performed using as inputs the parameters available  in possible experiments at the FLAME facility 
of the Laboratori Nazionali di Frascati: a gaussian modulating intensity with full width at half maximum (fwhm)
$l'\!\simeq\! 7.5 \mu$m (corresponding to a time $\tau'\!=\!25$fs), wavelength $\lambda\!\simeq\! 0.8 \mu$m, 
 $\E\!=\!5 $J,  
 $R$ tunable in the range $1\!\div\!10^{4}\mu$m; 
a supersonic helium jet or an aerogel  (if $\widetilde{n_{0}}(Z)\!=\!n_0\,\theta(Z)$ with $n_0\!\gtrsim\! 48\!\times\!10^{18}$cm$^{-3}$) as targets. The energy spectrum,  or equivalently the distribution $\nu(\gamma_f)$  of   the expelled electrons vs. their final relativistic factor, depends substantially on $\widetilde{n_0},R$; pleasantly, in the case $\widetilde{n_{0}}(Z)\!=\!n_0\,\theta(Z)\tanh(Z/a)$ it is peaked (almost monochromatic) around $\gamma_{{\scriptscriptstyle M}}$, the maximal $\gamma_f$.

\begin{table}
\begin{tabular}{|c|}
\hline
$\!$pulse energy ${\cal E}\!\simeq\!5$J, wavelength $\lambda\!\simeq\!.8\mu$m, 
 fwhm $l'\!\simeq\!7.5\mu$m, duration $\tau'\!\!=\!25$fs$\!$ \\
\hline
\begin{tabular}{|l|c|c|c|c|c|c|c|c|c|c|}
\hline
type of target    &  hj &  hj & hj &  hj & hj  &   ag & ag\\[2pt]
pulse spot radius $R \, (\mu$m)    &  16 &  8 & 4 &  2 & 2 &  2 & 1\\[2pt]
average intensity $I\,$(10$^{19}$ W/cm$^2$)  & 1& 4 &16 & 64 & 64 & 64 & 255\\[2pt]
asymptotic  density $n_0(10^{19}$cm$^{-3}$)    & 0.8  & 2   &  13   &   80   &   20  &   12 & 40 \\[2pt]
maximal relativistic factor $\gamma_{{\scriptscriptstyle M}}$   & 2.6 & 6 & 8.5  &  14 & 21 &   12.4 & 22.6\\[2pt]
maximal expulsion energy(MeV)     & 1.3  &  3  & 4.4  & 7.2 & 11 &   6.4 & 11.5 \\[2pt]
\hline
\end{tabular}
\end{tabular}
\caption{Sample inputs and corresponding outputs  if the target is: a supersonic helium jet (hj)
 or an aerogel (ag) with resp. continuous and step-shaped initial densities profiles as in fig. \ref{graphsb}. 
The expelled electron charge
is in all cases a few $10^{-10}$C 
}
 \label{tab1g}  
\end{table}

\medskip
On the other hand, for  $Z$ so large that $z_e(t,Z)$ keeps positive 
the map $\bx_e:\bX\!\mapsto\!\bx $ is again invertible and the solution (\ref{sol}) can be considered reliable 
 (by causality) for small $\rho/R$ and for $ct\!-\!z$ bounded by few $cT\!_{\scriptscriptstyle H}$. 
In particular, if  $\widetilde{n_{0}}(\!Z\!)\!=\!n_0\theta(\!Z\!)$ then by (\ref{expln_e}),  (\ref{sol'}) 
 well inside the plasma $n_e(t,z)$ is a travelling-wave 
with periodic peaks following the laser pulse, see  fig. \ref{graphsb} down-right. This describes the {\it plasma wakefield} 
in the {\it plane wave}  idealization \cite{Fio16}; as said, this can be considered reliable only for small $\rho/R$
and $ct\!-\!z$ bounded by few $cT\!_{\scriptscriptstyle H}$.

Summing up, we have proposed: 1. a new  laser-induced ``slingshot"  acceleration mechanism, 
which should yield well-collimated bunches of electrons of energies up to few tens MeV and  
is easily testable with present equipments; 2. 
a MFD description of the plasma wakefield just behind the laser pulse.

\end{document}